\documentstyle[preprint,aps,epsfig]{revtex}

\begin{document}
\draft
\title{The effect of atmospheric attenuation on inclined cosmic ray  air showers}
\author{M. T. Dova, L. N. Epele and A. G. Mariazzi\footnote{Correspondence: mariazzi@fisica.unlp.edu.ar - te: +54 221 4246062 - fax: +54 221 4252006}}
 \address{Physics Department, Universidad Nacional de La
Plata - C.C. 67 - 1900 La Plata, Argentina}
     
\maketitle

\begin{abstract}

The increasing cosmic ray statistics collected by present experiments and 
the future prospects with new large 
arrays demand accurate calculations of the extensive 
air shower parameters (EAS). The energy of the primary  particle
is estimated by ground arrays fitting a lateral distribution function (LDF)
to the particle densities at a given observing level. However, the 
lack of appropriate parameterization for these distributions, able to 
reproduce the 
data collected from all arrival directions, makes it difficult the
 experimental analysis. 
We propose a method to parametrize particle 
density distributions of EAS at any incident zenith angle.
Starting from analytical LDF for vertical showers we present a detailed study 
of the atmospheric depth dependence 
of the shower parameters. The results obtained are used to 
calculate the corresponding LDF for non-vertical showers including for 
the first time both, geometrical and atmospheric attenuation effects.
We check the method analysing electron and muon LDF generated 
by Monte Carlo simulations from incident cosmic ray particles at different 
zenith angles. A comparison of the proposed LDF with 
experimental results as well as MC data including detector effects is 
also presented.

\end{abstract}


\narrowtext

\section{Introduction}

The observation of cosmic ray particles  with energies 
well beyond the predicted Greisen-Zatsepin-Kuzmin cutoff \cite{agasa,hp,yak,agasa2},
enhanced by the discrepancies in the cosmic ray spectrum recently 
reported by HiRes and AGASA experiment \cite{agasa2,hiresicrc} 
is pushing astroparticle physics into a period of fast development 
both theoretically and experimentally.

The analysis of the extensive air showers (EAS) produced when the high 
energy particle enters in the atmosphere and interacts with the air 
molecules is crucial to know the characteristics of the ultra high 
energy cosmic rays (UHECR).
On the other hand, studies of the high energy interaction taking 
place in the atmosphere could yield unique information for elementary 
particle physics. 

The interest in EAS studies has been revived mainly because improvements
 achieved with new large arrays and the increasing statistics collected 
demand accurate calculations of the shower parameters. A very important 
shower observable is the spatial distribution of particles at a given 
 observing level because ground array techniques use the particle 
density distributions of EAS to 
estimate the energy of the primary particle, fitting a lateral 
distribution function (LDF) to the observed particle densities 
\cite{watsonagano} . 
There exist several LDF parameterizations well suited to describe 
density distributions  for primary incident angles smaller than 45 degrees, which are currently used by cosmic ray experiments.

High zenith angle showers induced by protons, heavy nuclei or photons have 
been subject of recent studies as an important tool to obtain relevant 
information about CR mass composition\cite{zas,prlzas}. 
Very inclined showers are also of great importance since they are the main 
background for neutrino detection as expected in the Pierre Auger 
Observatory \cite{desreport,capelle,xavierb}
Although CR experiments have detected showers with large zenith 
angles, most of these data 
have not been yet analyzed due to the difficulties present in understanding 
the space-time structure of the 
inclined showers. It is worth mentioning that the 
event reconstruction is performed in all air 
shower arrays 
assuming that the observation depends only on perpendicular distance 
from the shower axis. This very strong assumption, if not satisfied, 
may introduce important errors in core position and energy determination;
precisely this is the case for large zenith angle showers. The 
main difference in analyzing inclined showers lies in the fact that 
the assumed 
circular symmetry for the observed particle densities from vertical showers is 
broken. Besides, there is an asymmetry due to the 
different path traveled by particles in the upper and lower sides of 
the plane perpendicular to the shower axis (attenuation effect). 
These effects which are small 
at $\theta < 45 ^{\circ}$ become important at larger zenith angles, 
producing a significant asymmetry of the lateral distribution of particle 
density. On the other hand, the charged particle trajectories 
through the atmosphere are 
distorted in the geomagnetic field, effect extensively 
studied \cite{hillas,hp,andrews,antonov,ivanov}. 
A recent analysis of the influence of the magnetic field of the Earth on the 
shower observables for the Southern site for the Pierre Auger Observatory 
(average Geomagnetic Field of 25000nT and site altitude at 1400 m.a.s.l.)
\cite{cillis}
has been performed. The study
indicated that deflections alter significantly the lateral distribution of 
muons only at zenith angles $\ge 75^{\circ}$. This result, however, 
would be different if one considers different geomagnetic fields and 
ground level altitudes \cite{prlzas}. In all cases, for 
smaller angles, the asymmetry in the particle 
densities is dominated by both a purely geometrical effect and the effect 
on the longitudinal development. As a result, the azimuthal 
symmetry of the spatial distribution is destroyed, changing the 
dependence on the distance from the shower core, directly affecting the 
estimate of the energy of the primary particle. Much of the 
interpretation of EAS phenomenon involves the LDF; in addition to 
energy determination, the  
parameter characterizing the actual stage of the EAS development 
(the so-called age parameter) is expected to contain information 
about the primary composition as well as the characteristic of high energy 
interaction.

In this paper, we present an approach to analyze lateral 
distributions of particles produced by UHECR, which is valid 
at all 
zenith angles. We tackle the problem studying, by means of Monte Carlo 
simulations, 
the evolution with 
atmospheric depth of 
the parameters of semi-analytical lateral structure  
functions for vertical showers.  
The results obtained are then used to calculate the corresponding 
LDF for inclined showers including for the first time both, geometrical and attenuation effects.
We also present a comparison of the proposed LDFs 
with AGASA and Haverah Park experimental data and with MC data 
including water \v Cerenkov detector effects.

\section{Analysis of vertical showers} 
\subsection{Electron lateral distribution function}
We start our analysis using the  
lateral structure function derived for a pure electromagnetic cascade
by Nishimura and Kamata, and later worked out by Greisen 
\cite{nkg,nk} in the well-known NKG formula. This 
function has been much used in theoretical developments 
(3D extensive air shower simulation) as well as in the description of 
experimental results \cite{kascade,akeno}. The NKG formula
 is given by:

\begin{eqnarray}
\rho_{}(r) & = & \frac{N_e}{r_m^2} \frac{\Gamma (4.5-s_{nkg})}{2\pi\Gamma(s_{nkg})\,\Gamma(4.5-2s_{nkg})}\left( \frac{r}{r_m}\right)^{s_{nkg}-2} \left(1+\frac{r}{r_m}\right) ^{s_{nkg}-4.5}
\end{eqnarray}
with $
s_{nkg} = 3/(1+\frac{2 \ln(\frac{E_0}{\epsilon_0})}{t})$ and $
t = \int_{z}^{\infty}\rho_{atm}(z)dz/X_{0}$, where 
$N_e$ is the total number of electrons,
$\epsilon_0$ is the critical energy ,
$E_0$ is the primary energy,
$r_m$ is the Moli\`ere radius and
$X_0$ is the radiation length in air.
The age parameter, $s_{nkg}$,  characterizes the actual stage 
of the EAS development.

To obtain the lateral distribution of particles it is necessary to 
consider the multiple Coulomb scattering, which determines the characteristic 
size of the shower front. The lateral development in 
electromagnetic cascades in different materials 
scales well with the Moli\`ere radius \cite{pdg} given by $
r_m = E_s\frac{X_0}{\epsilon_0}$, with the energy scale 
 $E_s=m.c^2(\frac{4\pi}{\alpha})^{0.5}$.

It is straightforward to see that the Moli\`ere radius varies
with the inverse of the density of the medium.

\begin{eqnarray} 
r_ m & = &r_m(h_0)\frac{\rho_{atm}(h_0)}{\rho_{atm}(h)} = 
\frac{9.6g cm^{-2}}{\rho_{atm}(h) } 
\end{eqnarray}

In order to study the development of the shower parameters with $t$, we 
use Monte Carlo simulations of vertical proton induced 
showers at $10^{19}$ eV 
generated 
using {\sc aires}. This program is a realistic
 air shower simulation system, which includes electromagnetic algorithms 
\cite{hillas2}
 and links to different hadronic interactions models. We use in this 
paper {\sc qgsjet} \cite{qgsjet} for nuclear fragmentation and 
inelastic collision which is consistent with experimental data 
\cite{kascatere}. 
For the highest energy air showers the number of 
secondaries become so large ( $> 10^{11}$) that it is prohibitive, 
in computing time
and disk space, to follow and store all of them. A. M. Hillas \cite{hillasthi}
introduced a non uniform statistical sampling mechanism which
allows to reconstruct
 the whole EAS from a small representative fraction of secondaries
that are fully tracked. Statistical weights are assigned to the sampled
 particles to account for the energy of the discarded particles. This
technique is called ``statistical thinning''. {\sc aires} includes an extended 
thinning algorithm, which is explained in detail in reference \cite{aires}. 
The present work has been performed using in most cases an 
effective thinning level $\epsilon_{th} = E_{th}/E_{prim} = 10^{-8}$ 
which allows to avoid unwanted fluctuations and provide a 
reliable number of particles far from the shower core. All shower 
particles with energies above the following threshold were tracked: 
90 keV for $\gamma$'s, 90 keV for electrons and positrons, 10 MeV 
for muons, 60 MeV for mesons and 120 MeV for nucleons and nuclei.

With the particle data recorded we have evaluated the 
lateral distribution of electrons and positrons not only at ground altitude, 
but also at different 
atmospheric depths $t$. Fits to lateral distribution functions (LDF) 
were performed using equation (1) in order to obtain 
the corresponding parameters as a function of $t$, in a shower-to-shower
analysis. We found that the Moli\`ere radius  $r_m$ deviates 
from (2) when using a 
value of the age parameter fixed to the theoretical prediction value, which is an approximation valid  
for pure electromagnetic cascades. Undoubtedly, a different 
age 
parameter should somehow be used to reproduce showers initiated by hadrons. This point has already been addressed experimentally 
\cite{linsley,aguirre,porter,kawaguchi,akeno2,t5,t6,tt}
and extensively studied by several authors(See\cite{tt} and 
references therein). 

Our studies of vertical showers confirm, as proposed by \cite{capde}, that 
a modified NKG formula should be used.
Modifications of the NKG formula have been presented using theoretical 
considerations \cite{uchaikin} as well as parameterizations induced by
 Monte Carlo 
simulations . In the first case, the method is subject  to 
several approximations in order to obtain an analytical expression. 
The second one relies on Monte Carlo results at ground level 
yielding to average 
values of the parameters which are then included in the parameterizations. For 
our purpose 
it is important to obtain a very general LDF which can 
be used to describe the behavior of lateral distributions of electrons 
and positrons for vertical showers, not only at ground level but at 
any atmospheric depth. We use a NKG function with an age parameter given by
\begin{eqnarray}   
s & = \frac{3}{1+ 2 \,\beta\,/t}
\end{eqnarray}
where 
$\beta$ takes into account the above mentioned deviations from the 
theoretical value $s_{nkg}$.
The proposed NKG formula provides a good 
description of 
the LDF at all stages of shower development only outside the hadronic
core which lies close to the shower axis.  It should be mentioned
at this point that most UHECR experiments can only measure densities at 
$r> 100 $m from the axis, since in the very dense region near 
the shower core the particle numbers are so large that array detectors 
would be saturated for high energy showers.
Fits to 
Monte Carlo data at $r> 100 $m from the shower core were performed 
with $r_m$ fixed to (2), leaving the overall 
normalization and  $\beta $ as  free parameters.

Figure 1 (top) shows the Monte Carlo $e^+ e^- $ density 
distributions  corresponding to 
a single $10^{19}$ eV proton shower at selected  atmospheric depths 
from the whole set of levels considered. The error bars are included in all cases in the points. 
Solid lines are the results of the 
fits. The reduced $\chi ^2$ are in the range 
1-3 with better fits around the position of the shower maximum.

The behavior of the fitted age parameter with atmospheric depth is
displayed in Figure 2. A set of 15 proton initiated showers was used to 
analyze the natural fluctuations.  With different symbols it is 
shown the average fitted $s$ value at different distances from the shower 
core; 150 m $<$ r $<$ 450 m (circles),  150 m $<$ r $<$ 750 m 
(squares) and 150 m $<$ r $<$ 1600 m (triangles). 
The 
error bars, in most cases smaller 
than the points,  represent the natural shower-to-shower fluctuations (The RMS fluctuations of the mean are always smaller than the symbols). 
A remarkable feature is the fact that 
the whole radial range beyond $150 $m can 
be described by a unique age parameter independent of $r$. 
The band 
in the same figure shows the age parameter calculated 
using (3) with the assumptions that $\beta $ is given by the 
position of average shower maximum, $<X_{max}> = 730 \pm 36 $g cm$^{-2}$ for 
the shower set used. It is evident that, around the shower maximum 
the agreement is better and the value of the age parameter is approximately 
1 at $X_{max}$.

The total number of electrons obtained from the fit to each 
 single shower 
is slightly lower 
than the actual one. This can be explained due to the fact that 
the shower size depends on 
extrapolation of the LDF in the region close to the core, which 
is not included in the fit. 

\subsection{Muon lateral distribution function}

It is well established that at ground level and 
for large zenith angles the muonic component
of an EAS becomes dominant \cite{cillis,zas}, reaching a 
maximum proportion at $75^{\circ}$, after which it decreases slightly 
\cite{cillis}. Consequently, the study of muon density distributions at all 
zenith angles is very important. Several parameterizations 
have been presented
for the muon lateral distributions produced in  vertical showers; 
a NKG-type LDF was empirically 
derive by Greisen\cite{nkg}

\begin{eqnarray}
\rho_{\mu}(r)& = & N_{\mu}(t) f_{\mu}(r) \nonumber \\ 
f_{\mu}(r,s)   & \approx &  \left( \frac{r}{r_G} \right)^{-0.75} \left( 
1+\frac{r}{r_G} \right)^{-2.5} \nonumber 
\end{eqnarray}
fixing $r_G=320m$. Vernov et al(\cite{vernov} and references therein) 
proposed a semi-analytical form of the structure function

\begin{eqnarray}  
f_{\mu}(r)   & = & C \,\left(\frac{r}{r_0}\right)^{-\gamma}\exp \left(-\frac{r}{r_0}\right)
\end{eqnarray}
with $\gamma=0.4$ and $r_0=80$m.
A similar approach was suggested by Hillas \cite{hillas}.
The slope of the two functions are in very good agreement at intermediate 
distance, but Vernov distribution is flatter close to the shower core and 
decreases faster at larger distances. These LDF have been used to fit 
experimental data, however, neither function seems to reproduce
the whole radial range of an EAS. The reason for that is clearly the lack of 
dependence on shower stage development, using instead fixed parameters 
in the corresponding structure functions. Very 
recently, the KASCADE experiment has used a NKG formula to fit muon density 
distributions. The fits were performed close to the shower core ($r<200$m) 
with non conventional $r_m$ and $s$ values\cite{kascade}.

A dependence of the parameters according to the shower age must be 
expected for the LDF although not precisely the same as the one exhibited by
electromagnetic component.
Unlike electrons, muons in an air shower are less attenuated and little 
affected by Coulomb scattering. Its spread is  determined by the direction 
of emission of the parent particle and hence increases while the shower 
propagates downwards. 
A shower age dependence of the muon structure function has been considered in 
relation with the parameter $\gamma$ (See \cite{alesio} and 
references therein).

For the purpose of this paper the Vernov function seems 
to be a better choice due to its 
analytical origin. In order to study the depth dependence of the parameters, we
 use a modified Vernov distribution 
with a $\gamma$ parameter 
dependence with atmospheric depth given by $\gamma =  2-s $, 
with $s$ as in (3)  due to the fact that in the radial region smaller than
the scale radius, 
 the slopes predicted by both the NKG-like 
and Vernov distribution are very similar.
  
It has been shown \cite{bosia,berga} that, if the parent particles are 
created with a $p_{\perp}$ distribution: $ p_{\perp}/
p_0 \exp(- p_{\perp}/p_0) dp_{\perp}/p_0 $, 
then the Vernov distribution is obtained at ground level, with $r_0$ given by
\begin{eqnarray}  
r_0 =   & = & \frac{2}{3}<H_p>\,\frac{<p^{\mu}_{\perp}>}{<E_{\mu}>},
\end{eqnarray}
where $<p_\perp>= 2p_0$ 
is the mean transverse momentum, $<E_{\mu}>$  the mean energy of muons
 and  $<H_p>$ the mean height of production. 

These approximate expressions can 
serve to calculate the variation with $t$ of the parameters characterizing 
the lateral spread.
The ratio $<p^{\mu}_{\perp}>/<E_{\mu}>$ can be considered constant while 
the shower develops \cite{zas} and the variation of $r_0$ with altitude 
is only determined by the $t$ dependence of the mean high of origin of muons. 
These particles are produced in every pion generation and their energy
 distribution follows that of their parents. Thus, one expects the
 dependence of 
 $<H_p>$ with $t$ to arise from two sources. The trivial one resulting from 
the fact that we consider observing levels at different altitudes,  
which gives a factor  $h_0\, \log(t/t_p)$ assuming an isothermal atmosphere 
of scale height
$h_0$; and a dynamical one due to the pion decay process. It is well 
known that, at a given energy, the pion decay probability 
decreases with increasing air density. Thus, a characteristic radius
$r_0(t)$ results,

\begin{eqnarray}  
r_0(t) =   & = & \frac{2}{3}\,
\frac{<p^{\mu}_{\perp}>}{<E_{\mu}>}\, h_0\, \frac{t_G}{t}\,\log(\frac{t}{t_p})
\end{eqnarray}

where $t_G$ is the ground level depth in $g cm^{-2}$. Note that at 
 $t = t_G$ equation (5) is recovered. 

Fits using the modified Vernov function to the muon density distributions obtained from the same
Monte Carlo set of showers generated as described in 
the previous section, were performed with $N_\mu $, 
$r_0$, and $\beta $ simultaneously free. Reduced $\chi ^2$ of the fits 
to single showers for all
predetermined observing levels are in the range 1-3. 
Figure 1 (bottom) displays the  $\mu^+ \mu^- $ 
density distributions (points) for a single $10^{19}$ eV proton shower 
corresponding to a 
subset of the atmospheric levels, together with the fits 
(solid lines) where the agreement is evident.

The parameter $\beta$ is approximately constant and much bigger 
than the shower $X_{max}$. This feature is somehow consistent with 
the fact that muons become relatively more and more abundant as the altitude 
decreases with only a slight attenuation.

The mean value of $r_0$ from the above mentioned set of 
showers is plotted in Figure 3, where error bars indicate the 
natural shower-to-shower fluctuations. The solid line 
corresponds to a fit using equation (6) which gives $t_p$ of the 
order of $\Lambda_N$, the attenuation length for nucleons in air,
as expected. Note
that $r_0$ is a flat function of $t$ resulting from a compensation of the 
two factors described above. For inclined showers, the typical slant depth 
of the shower increases significantly with the zenith angle and the $t$ 
dependence becomes dominated by the atmospheric density factor. 

Figure 4 displays the obtained age parameters 
fitting different radial ranges: 100-300 m (circles), 100-600m (squares) 
and 100-1600 m (triangles). The 
error bars indicate as in the other cases the natural shower fluctuations. 
 The curve shows  
 the prediction using  
(3) and constant value of  $\beta = 1800$ g cm$^{-2}$.
The total number of muons $N_{\mu}$ from the fits agrees quite well 
with the corresponding values registered from the Monte Carlo 
data, in spite of the fact that 
the fits are performed in all cases at core distances  $r>100 $m, implying an
 extrapolation of the LDF beyond the range considered.  

\section{Inclined showers}

The main goal of the analysis done on vertical shower was to obtain 
the atmospheric depth dependence of the shower parameters. This result is 
crucial to properly
take into account the effect of the attenuation in the atmosphere 
observed in inclined showers, as we shall show in what follows. 
In this section we extend the solutions obtained for vertical showers to 
inclined ones, considering not only the geometrical effect but also
 the 
influence of the attenuation in the atmosphere.

It must be emphasized that the measured data by surface 
arrays are ground density of particles and the showers are analysed
projecting the ground densities into the shower plane. As mentioned in 
the introduction, this procedure 
allows correction of the geometrical effects and it is a good approximation 
for near-vertical showers. However, it was shown in the section II for vertical showers, that the parameters of the LDF depend on 
the atmospheric depth $t$. This means that to explain the observed 
asymmetry, besides the transformation to shower 
front coordinates, one has to evaluate all parameters of the LDF at 
the new slant depth $t'$

\begin{eqnarray}   
t' & = & \frac{\int_{z'}^{\infty}\rho_{atm}(z)dz'}{X_0}
\end{eqnarray}
with $z'$ along the shower axis direction, in order to include the 
attenuation effect in the atmosphere. 

Actually, the structure of an air shower is quite complex but 
following \cite{pryke,dova}, an inverted cone seems to be 
a reasonable first approximation.
The intersection of an inclined cone with a plane parallel to the ground 
gives,
\begin{eqnarray}
\frac{z}{\cos\theta}-L & = & (z'- L)\,\left (1-\frac{ r'}{z'- L}
\tan\theta \cos\phi \right)
\end{eqnarray}

The parameters $L$, $\theta$ and $\phi$ are indicated in Figure 5. 
If the cone angle $\alpha $ is assumed nearly 
independent of $z$ then,
 the slant depth $t'$ results from: 
\begin{eqnarray}
\frac{t}{t'} & = & \cos\theta\,(1-\tan\alpha\,\tan\theta\,\cos\phi')
\end{eqnarray}

Particle density distributions for inclined showers are given by 
the corresponding electron and muon LDF but evaluated at slant 
depth $t'$ where the 
dependence on 
the azimuthal angle is evident. 
Proton air showers at $10^{19}$ eV were generated under the same conditions of vertical showers with 
{\sc aires} + {\sc qgsjet} at zenith angles between $\theta=10^{\circ}$ 
and $\theta=75^{\circ}$, where the influence of the geomagnetic field is 
still negligible. The geomagnetic field was then  switched off in the simulation. 
The primary particle arrives at the corresponding zenith angle from 0$^{\circ}$
East of the South.

For single showers, two-dimension density distributions $r$-$\phi $ were 
built at all predetermined observing levels for different 
particles with bins 100 m and  30 deg wide. 

\subsection{Electron lateral and azimuthal distributions}

The distributions were fitted using the modified NKG formula with $s$ (Eq.(3)) 
including the 
complete zenith angle dependence, with $\alpha$, $\beta$ and $N_e$ as free parameters.
The proposed LDF  fits very well $e^+e^-$ distributions at all 
observing levels 
for $\theta \le 60^{\circ}$. 
At zenith angles greater than $60^{\circ}$ the fits are good only up to a 
slant depth $t'\sim2000$g cm$^{-2}$. This can be understood at the light 
of the physics behind air showers 
induced by nucleons or nucleus. These particles, undergo nuclear 
collisions in the atmosphere. The cascade begins with a hadronic 
interaction with an increasing number of pions in each generation of 
particle interactions. The rapid decay of $\pi^0$'s into photons 
feeds the electromagnetic cascade, which finally 
dissipates $\approx 90 \%$ of the primary energy while charged pion decays 
produce the muons and neutrinos. The development of the shower 
reaches a maximum 
at around 800 g cm$^{-2}$ 
for vertical 
$10^{19}$ eV proton showers. However for inclined showers, the slant 
depth rises dramatically from $\approx$ 1000 ($\theta = 0^{\circ}$) to 
36000 g cm$^{-2}$ ($\theta = 90^\circ$) . A detailed study 
of the characteristics of the muon and electromagnetic component in very 
inclined showers can be found in reference \cite{zas}. 
A standard feature of very inclined showers is that particles at ground 
are dominated by the muonic component, while the electromagnetic component is 
almost completely 
absorbed due to the greatly enhanced atmospheric slant depth. Beyond 
$\approx$ 2000 g cm$^{-2}$ the small electromagnetic cascade is mostly 
originated 
from muon decay and one cannot expect a NKG distribution for electrons 
originated from muon decay.

Figure 6 shows a set of $r$ projection (integrated contents over all 
$\phi$) and $\phi$ projections ( integrated contents over the fitted $r$ range)
on the shower front from  the 
two-dimension distributions at given atmospheric depths, corresponding to 
$30^{\circ}$ and $60^{\circ}$ 
inclined 
showers. Points with error bars stand for
simulated data and curves represent the result of the fit.
It is important to note the  asymmetry due to atmospheric 
 attenuation effects, higher particle density at $0^{\circ}$ 
azimuth angle and a depression at  $180^{\circ}$. This is 
evident in the plots showing 
$\phi $ projections, even for the case  $\theta = 30^{\circ}$, becoming very important at larger zenith angles. Note 
also the symmetry in the density at $\pm 90^{\circ}$. It
can be seen from the plots that our approach for the LDF
 predicts the correct dependence for the azimuthal angle distribution. 
In observations with large zenith angles, the position 
of the depth of maximum is more distant from 
the ground with an increasing atmospheric absorption of 
the electromagnetic component. However, the plots indicate that, even at
$\theta < 70^{\circ}$
a significant number of electrons can reach the ground level.

Figure 7 (top) displays the fitted age parameters at selected atmospheric 
depths for single 
showers at  $0^{\circ}$ (circle), $30^{\circ}$(square), $50^{\circ}$ (triangle)
 and $60^{\circ}$ (diamond). When 
plotting together the 
results obtained from different showers at different 
levels, it is put into evidence that the fitted age parameters 
overlap smoothly
with only small differences due to shower-to-shower fluctuations.  
This result implies a consistency check and 
confirms the validity of our approach. 
With the same aim, the corresponding total number of electrons 
obtained from the AIRES output are plotted in solid lines in Figure 7 (bottom),
along with the fitted shower size $N_e$ (points with 
same convention than in Figure 7 (top)) at selected atmospheric depth. 
It is once more worth mentioning that the curves which correspond to single 
showers with  different incident zenith angles are of course, 
affected by the fluctuation of the first interaction point. However, 
the shower size obtained from the fit agrees rather well with the 
values from the Monte Carlo results, showing the expected longitudinal shower development.

The parameter $\tan(\alpha)$ takes values between 0.15 and 0.2, as
expected, since an EAS covers a large area but subtends a small apex angle.

\subsection{Muon lateral and azimuthal distributions}

The same set of generated showers, proton air showers at $10^{19}$ eV 
at zenith angles between $\theta=30^{\circ}$ 
and $\theta=70^{\circ}$ with an effective thinning level of $10^{-8}$,
was used to analyze the muon component. 

Using the procedure described in the previous section for electrons, 
for each single shower two-dimension density distributions $r$-$\phi $ 
were 
built at all predetermined observing levels. In order to perform the fit 
to the muon densities resulting from the inclined showers, the modified 
Vernov LDF proposed in section II.B was used. The azimuthal angle dependence 
was included following the approach described in section III. 
The proposed function
successfully  fits the Monte Carlo data at 
all zenith angles. Reduced   $\chi ^2$  are in the range $\approx 1-3$. As in the 
case of electrons, the fitted $N_{\mu}$ and $s$ values are compatible
 with the expected ones. 
Figures 8 show the radial and azimuthal angle projections of the 
muon density from two-dimensional fits to the data 
for $60^{\circ}$ and $70^{\circ}$ zenith
 angles. The dots stand for the Monte Carlo data and the curves 
correspond to the fits. It seems worth noting that the asymmetry 
introduced by the 
atmospheric attenuation in the lateral distribution is small, 
mainly at observing levels near 
the ground. This feature contrasts with the observed one
for electrons, even at lower zenith angles. In previous 
analysis of quasi-horizontal showers \cite{zas}, the attenuation 
effect was neglected. However, as mentioned in the 
introduction, for showers arriving at 
$\theta >75^{\circ}$ (with high $\mu $ content at ground level) the 
lateral asymmetry is dominated by geomagnetic effects. The complex 
behavior of the muon component due to magnetic field effects was 
studied in great detail in reference [7].

\section{Comparison with experimental LDF}

As mentioned above, to estimate the energy of the primary 
particle, analysis of the lateral distribution of shower particles at ground 
are important. Depending on the particle detectors of ground array 
experiments, different LDF 
parametrizations have been used. The signal measured by
plastic scintillators (Volcano Ranch, Yakutsk and AGASA)  is dominated 
by the electromagnetic component of the shower compared with the muon signal.
In Water \v{C}erenkov detectors (as used in Haverah Park and being implemented for 
the Pierre Auger Observatory) the electromagnetic and muon 
components contribute to the total signal. Figure 9 shows the signal produced by different particles (top) and the 
ratio of the muon to 
electromagnetic 
\v {C}erenkov signal (bottom) measured in units of vertical equivalent 
muons per m$^2$  for a $10^{19}$ eV vertical proton 
shower. It is evident from the plot that the signal changes with distance 
from the core, being 
dominated by the muon component far from the shower axis. 

The implementation of the method proposed in this paper to obtain the 
LDF for inclined showers once the appropriate  parametrization for vertical showers 
is found,  clearly implies the analysis of the parameters of the 
corresponding LDF to find their dependence with depth. This specific 
study has 
to be performed in all cases using simulations.  

In what follows, we compare the 
parametrizations proposed in section II and III for particle densities 
with experimental data. 
It should be stressed that we do not intent to re-analyse the experimental 
data but rather to investigate to which extent the obtained LDF from pure 
shower 
simulations are valid to reproduce experimental data without including 
additional modifications. In the case that the LDF are suitable to 
describe experimental data, it would be possible to further check 
the method presented in section III.
  
For scintillator detectors, generalizations of the NKG distribution 
have been used to parametrize the observed density with distance. 
In Figure 10 we present the experimental result of AGASA 
for vertical $10^{19}$ eV showers
\cite{nagano} in the form of 
their empirical functions for the total signal S(r) and density of muons of 
threshold energy 1 GeV,  together with 
the result of the fit using the parametrizations proposed in this paper. 
The 
S(r) signal was fitted with the modified NKG function of section II, 
with  the conventional 
assumption of $r_m = 91.6$ m. This function can reproduce 
the AGASA experimental LDF over the core distance 100-2500 m surprisingly 
well with a unique value of the age parameter $s = $ 1.05. If one assumes 
a pure electromagnetic cascade, it is possible to make a very crude 
estimation
of the primary energy from the fitted $\beta $ parameter which 
results to be $E = 1.7 \times 10^{19}$eV.
The value of the signal density at 600 m 
from the core with our fit is 41 m$^{-2}$ to be compare with the one 
obtained from the AGASA results $S(600) = $51 m$^{-2}$. The average 
conversion relation for primary proton and iron using two different 
hadronic interaction models  determined by the AGASA 
experiment is $E[eV] = 2.23\times10^{17} S_{0}(600)^{1.02}$ \cite{sakaki}. 
Using this expression we can estimate the energy from our fit to be
 $E = 1.2 \times 10^{19}$eV.
  
In addition, we fitted the AGASA muon lateral distribution with 
the modified Vernov function from section II.B
in the range 100-2500 m with all parameters left free. The result of the fit 
yields to $r_0 = 436$ m. It can be seen that the shape of 
the fitted LDF is consistent with the empirical AGASA function.  The value 
of $\rho_\mu (600)$ from 
our parameterization is 3.86 m$^{-2}$ in perfect agreement with the 
value obtained using the AGASA function. 

Due to the lack of published signal densities in the X-Y plane for 
large zenith angle showers from AGASA experiment, we present here 
only comparisons with 
vertical showers. It would be straightforward to 
extend the analysis to non-vertical showers 
using the same LDF and the method we 
proposed in section III. It is worthwhile to remark that most of 
the experiments have only published cosmic ray showers at small 
zenith angles.

Water \v{C}erenkov signal (WCS) due to EAS particles at ground includes 
both electromagnetic particles and muons, and the signal  produced 
in the detectors is the sum of both components. In Figure 11 we show an 
extremely large EAS observed by the Haverah Park experiment and their
 predicted LDF in the regions $r<800 $m and   $r>800 $m as 
described in \cite{tt} (solid lines). The result of the fit using 
the modified NKG formula from section II.A is also plotted. 
The Moli\`ere radius 
was fixed to the value calculated using (2) and the mean atmospheric 
depth of the array; 1018 g cm$^{-2}$. Our LDF reproduces rather well the 
Haverah Park data in the whole range with age parameter $s = 1.24$.
From the fitted function a value 
of $\rho$(600) =  23.8 VEM m$^{-2}$ was obtained to be compared with the value
obtained by Haverah Park  $\rho$(600) =  25.2 VEM m$^{-2}$.

In order to check the 
validity of our approach for Water \v{C}erenkov detectors we have also made 
estimations of the WCS using a direct conversion procedure that 
retrieves average signals for every particle hitting the 
detector \cite{sergio}. Such averages were evaluated using the AGASIM 
program \cite{clem}. Fig. 12
(Top) shows the obtained WCS for all the particles at ground level 
for a $10^{19}$ eV vertical shower for distances to the 
axis beyond 200 m (dots). The muon 
threshold energy is 300 MeV. 
The result of the fit using the modified NKG formula leaving all 
parameters free is superimposed as solid lines, providing a good 
description of the data 
($\chi2 /d.o.f = 26/12 \approx 2$). The scale radius 
determined by the fit is 1.4 times 
the pure theoretical Moli\`ere radius. The 
age parameter results $s=1.15$, being strongly correlated with $r_0$. 
From the fitted function a value 
of $\rho$(600) =  12.7 VEM m$^{-2}$ was obtained. With this value and the 
  energy calibration formula used to analyse the Haverah Park 
data \cite{hpicrc} ($E(EeV) = 0.612 \times \rho(600)^{0.99}$), 
the estimated primary energy results  $E_{p} \approx 0.8 \times 10^{19}$eV. 
Using 
the parametrization for non-vertical showers derived 
by application of our method to 
include the effect of the attenuation in the atmosphere in the 
modified NKG function, we also 
fitted the lateral density of WCS at ground for a proton shower at zenith 
angle of $40^{\circ}$. At this modest zenith angles the average 
signals estimated as described above
 are still the appropriate ones. The resulting LDF fits well the data  
($\chi2 /d.o.f = 383/165 = 2.3 $) with a fixed value of 
$r_0 = 1.4 r_m$ as for the vertical showers. The s parameter tends 
to decrease. It is, however, well known that the shower 
parameters $s$ and $r_0$ are highly correlated.
The fitted parameter $\tan(\alpha)$ takes a value of 0.1 in agreement with the expected one. In Figure 12 (bottom) the radial and azimuthal 
WCS density projections, measured in VEM m$^{-2}$, are plotted 
together with the result of the fitted LDF.

\section{Conclusions}

It is clear that information concerning the lateral 
distribution functions of particles is very important to 
determine the characteristics of the ultra high energy cosmic rays. 
In this paper, we presented a new approach to obtain appropriate 
parameterizations of LDF, able to reproduce particle density 
distributions generated for UHECR arriving from any zenith angles.
 
We started our study with the analysis of  vertical showers using
modified NKG (section II.A) and 
Vernov (section II.B) structure functions for electrons and muons,
 respectively. 
A detailed study of the shower parameters with atmospheric depths was 
performed. The results  
indicate that a unique age parameter can be used at radial distances 
greater than 150 m from the shower core, which is the region of
 interest for surface arrays. The knowledge of the behavior of shower  
parameters as the cascade of particles develops is crucial to 
work out the LDF 
for inclined showers. 
It is worth noticing that the method presented here it is well suitable to 
a shower-to-shower based analysis. 
We paid special attention to the 
fact that the LDF of particles from inclined cascades must be corrected
not only due to geometrical effects but also for the longitudinal attenuation. 
This latest, usually neglected or considered in a very simple manner 
in the experimental analysis, is surprisingly large for 
electron distributions, 
even at moderate zenith angles where the amount of electrons 
reaching ground level dominates. Thus, data analysis assuming that 
the lateral distributions depend only on the distance from the shower
 axis might yield to biased results if the atmospheric 
attenuation effect is not 
considered at all. 
Muon distributions are also affected by attenuation effects, although 
to a less 
extent. The asymmetry at very large zenith angles in this case
is dominated by geomagnetic effects. 

The obtained lateral distribution functions fit very well the generated 
data at all observing levels along the longitudinal shower paths. The very
good agreement of the theoretical distributions for inclined showers 
obtained using our method with the radial, and more important, the 
azimuthal angle distributions, provides strong supporting evidence 
for the validity of this approach. In connection with this result we 
have checked our method with the signal detected by ground array 
detectors. We have shown that if the signal produced in 
water \v{C}erenkov tanks is  parametrized by a modified NKG function, then  
the application of our method is straightforward and the corresponding 
LDF for inclined showers reproduces the experimental density distributions 
rather well. 

A final comment on the method proposed is well worth making here: 
Several present experiments use both NKG and Vernov type LDF. 
The application of our method would allow them to include high 
zenith angle showers in their analysis,  what implies a 
significant increase of their acceptance. Future experiments will 
also benefit from the analysis of showers arriving from any direction

\begin{figure}
\begin{center}
\epsfig{file=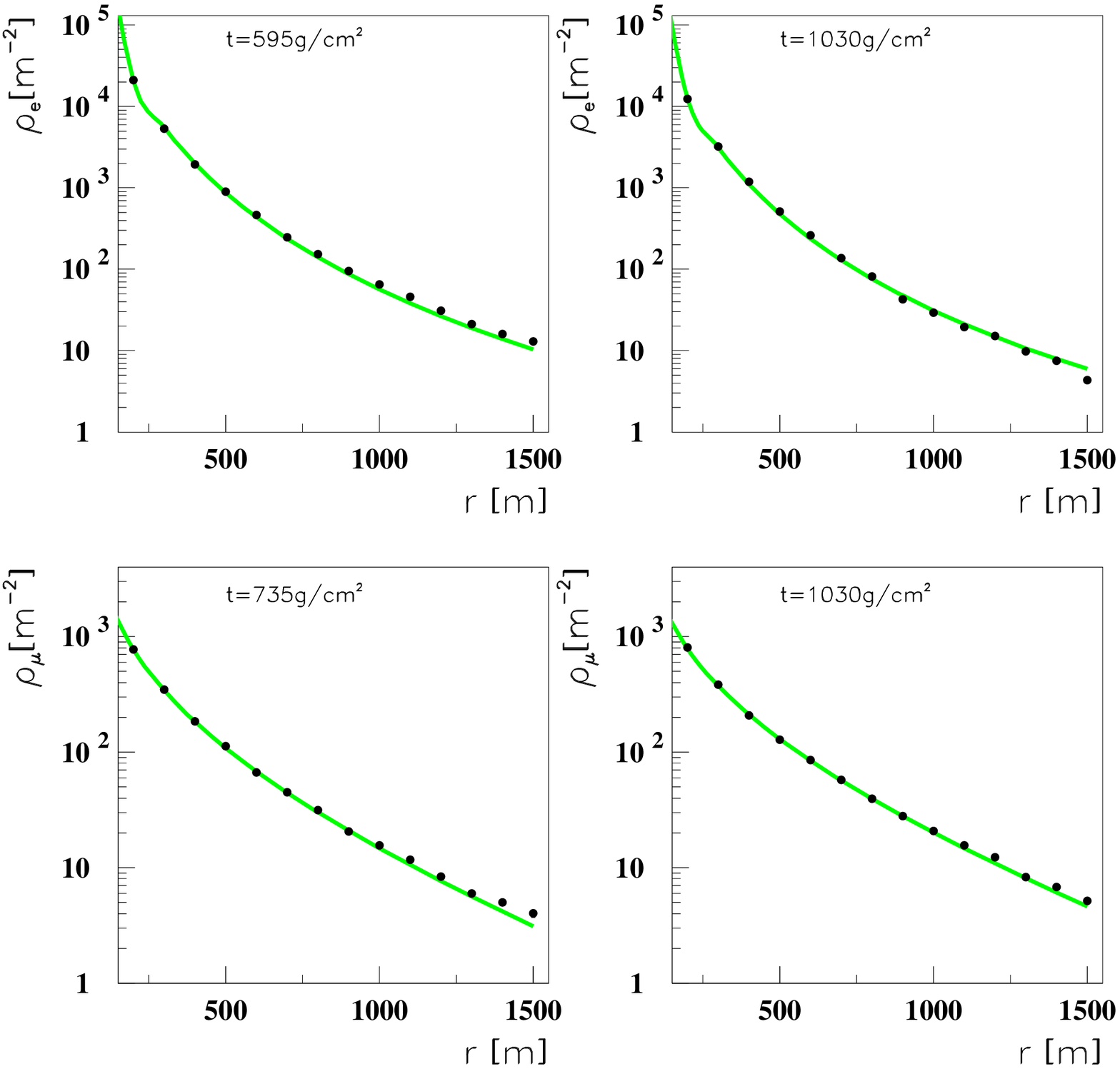,width=13.cm,clip=}
\caption{}
\end{center}
\end{figure}

Fig. 1: Electron (top) and muon (bottom) lateral distributions 
of a $10^{19}$ eV vertical proton shower at different atmospheric 
altitudes. The solid line indicates the result of the fit 
using the proposed LDF (See text).

\newpage

\begin{figure}
\begin{center}
\epsfig{file=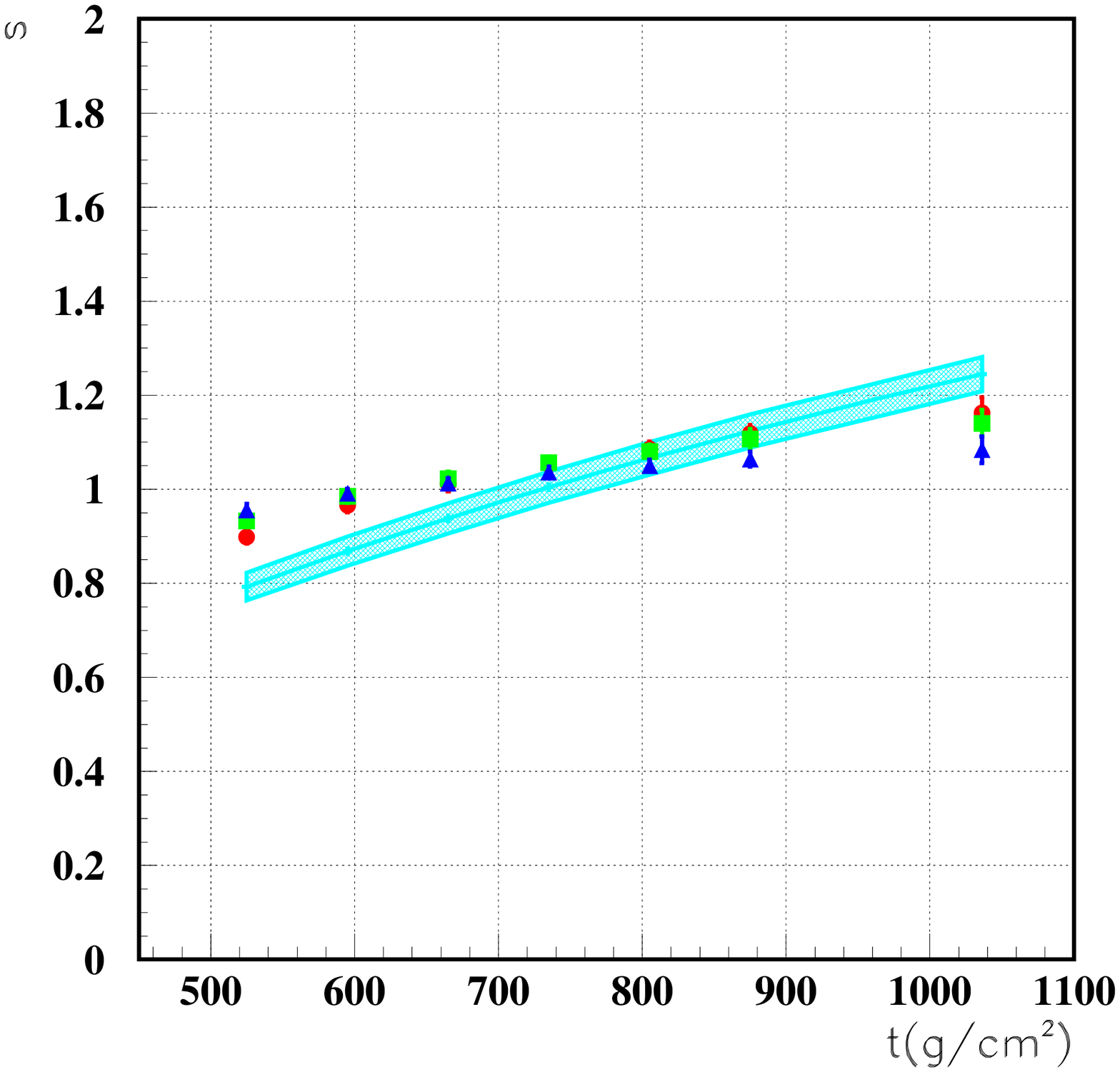,width=11.cm,clip=}
\caption{}
\end{center}
\end{figure}

Fig. 2: Atmospheric depth dependence of the average age parameter 
obtained fitting electron lateral distributions of single 
showers at different radial distances: 
150 m $<$ r $<$ 450 m (circles), 150 m $<$ r $<$ 750 m (squares) and 
150 m $<$ r $<$ 1600 m (triangles). 
Shaded area indicates the expected dependence using a NKG formula with (1).

\newpage

\begin{figure}
\begin{center}
\epsfig{file=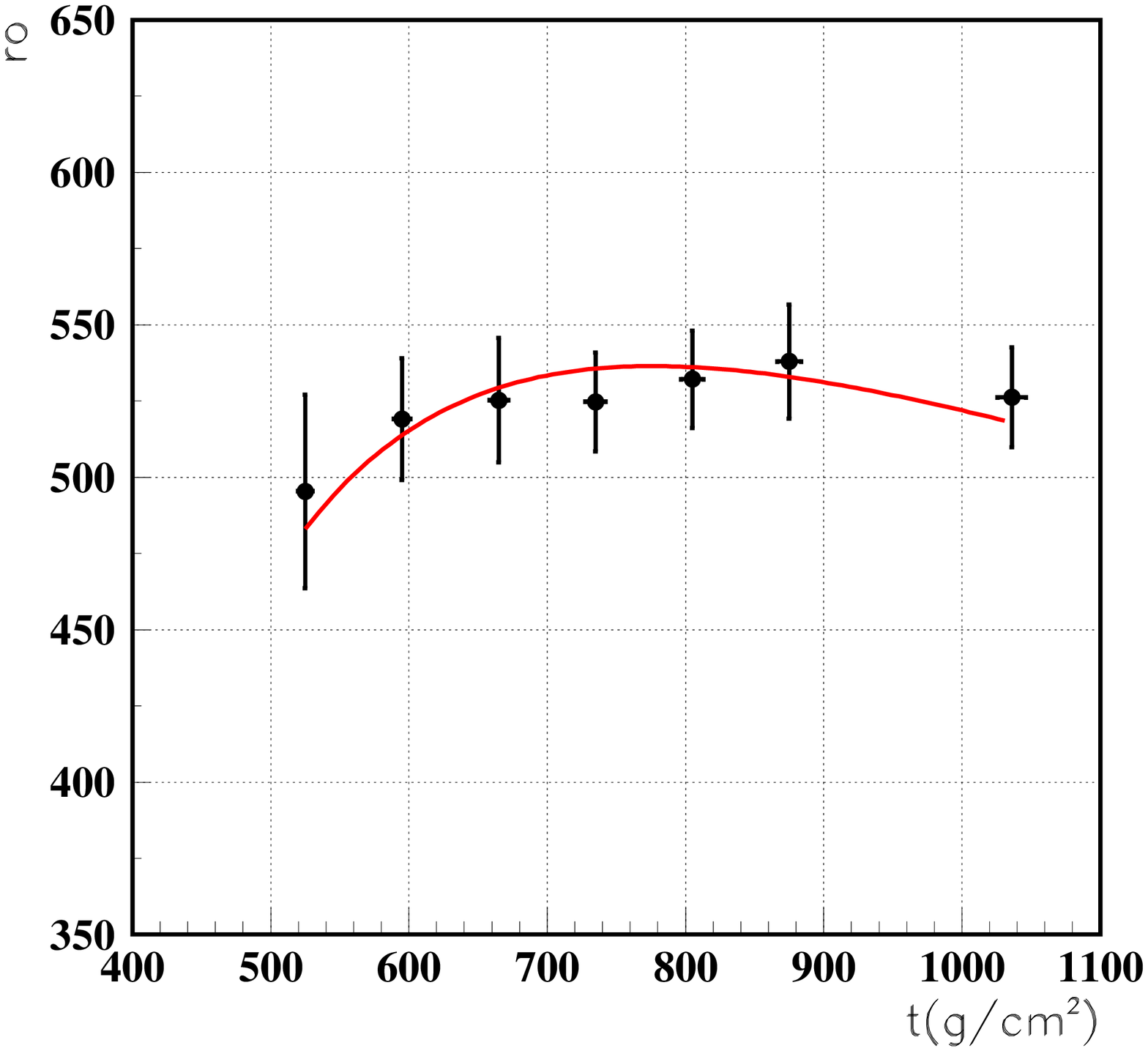,width=11.cm,clip=}
\caption{}
\end{center}
\end{figure}

Fig. 3: Atmospheric depth dependence of the average scale radius $r_{0}(t)$ fitting single showers. The error bars correspond to the standard fluctuations. Solid 
line is the result of the fit using equation (6).

\newpage

\begin{figure}
\begin{center}
\epsfig{file=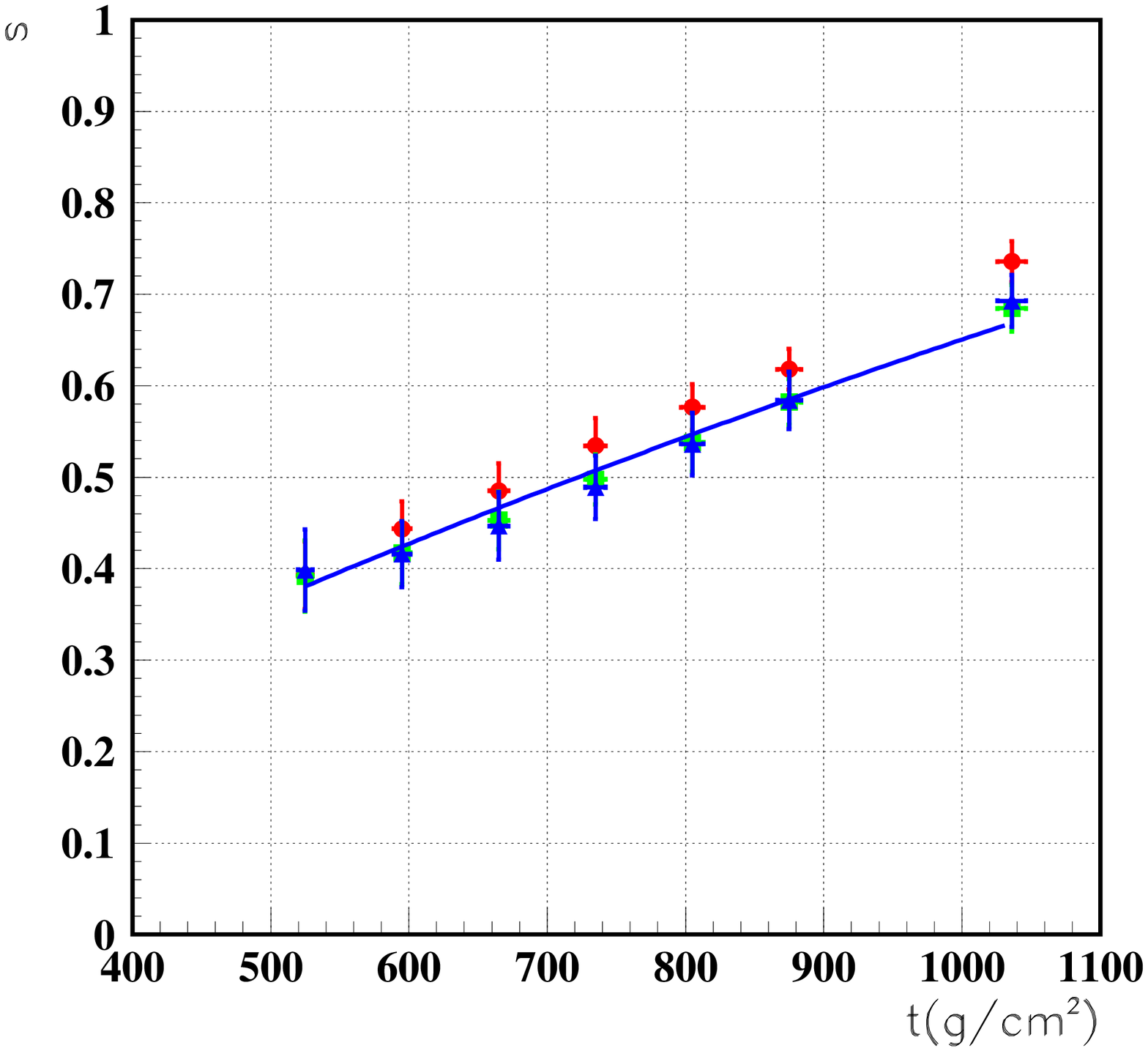,width=11.cm,clip=}
\caption{}
\end{center}
\end{figure}

Fig. 4: Atmospheric depth dependence of the average age parameter 
obtained fitting muon lateral distributions of single 
showers at different radial distances: 
100 m $< $r $<$ 300 m (circles), 100 m $<$ r $<$ 600 m (squares) and  
100 m $<$ r $<$ 1600 m (triangles)
The line indicates the fit as described in the text.

\newpage

\begin{figure}
\begin{center}
\epsfig{file=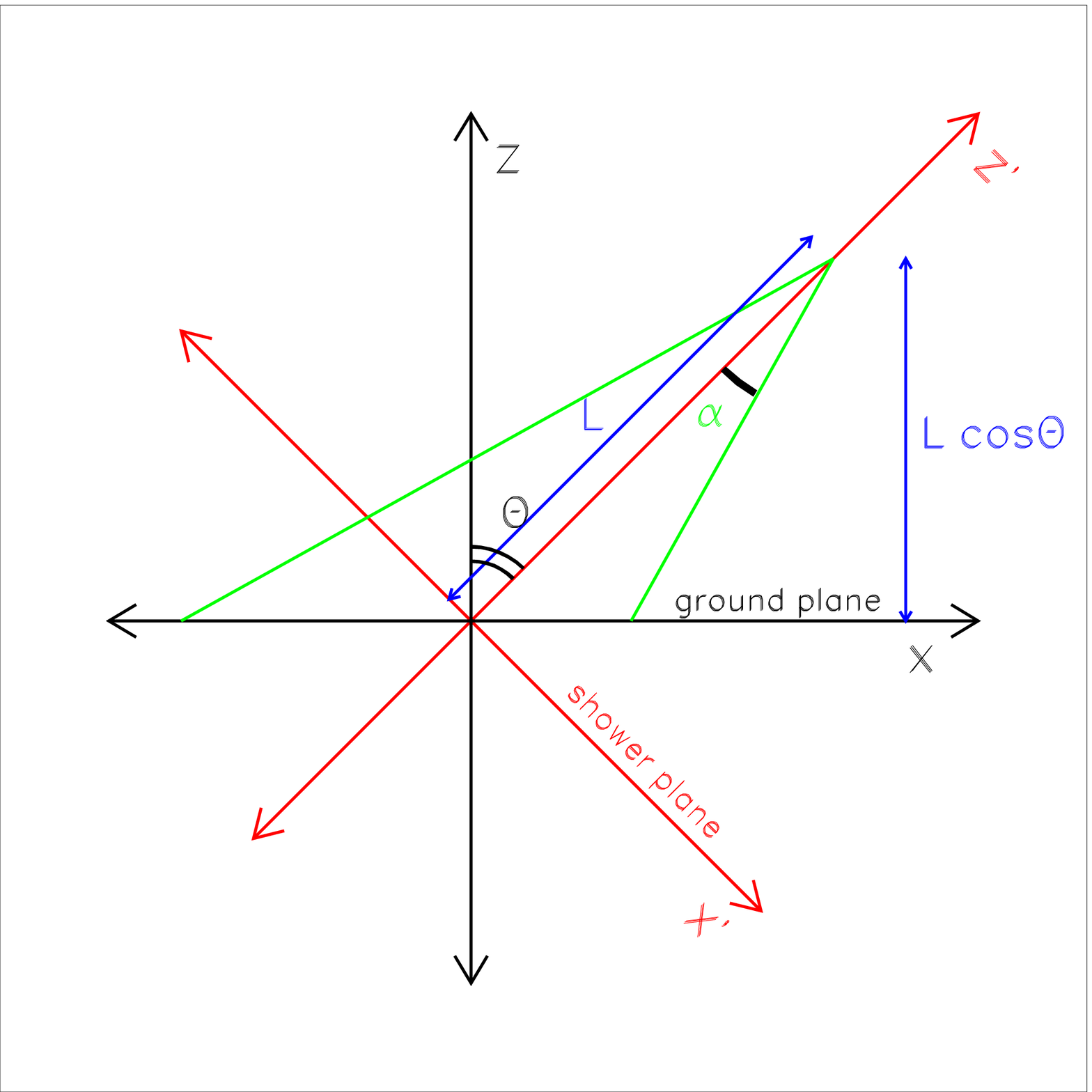,width=13.cm,clip=}
\caption{}
\end{center}
\end{figure}

Fig. 5: The intersection of an inclined cone and the ground plane.

\newpage
\begin{figure}
\begin{center}
\epsfig{file=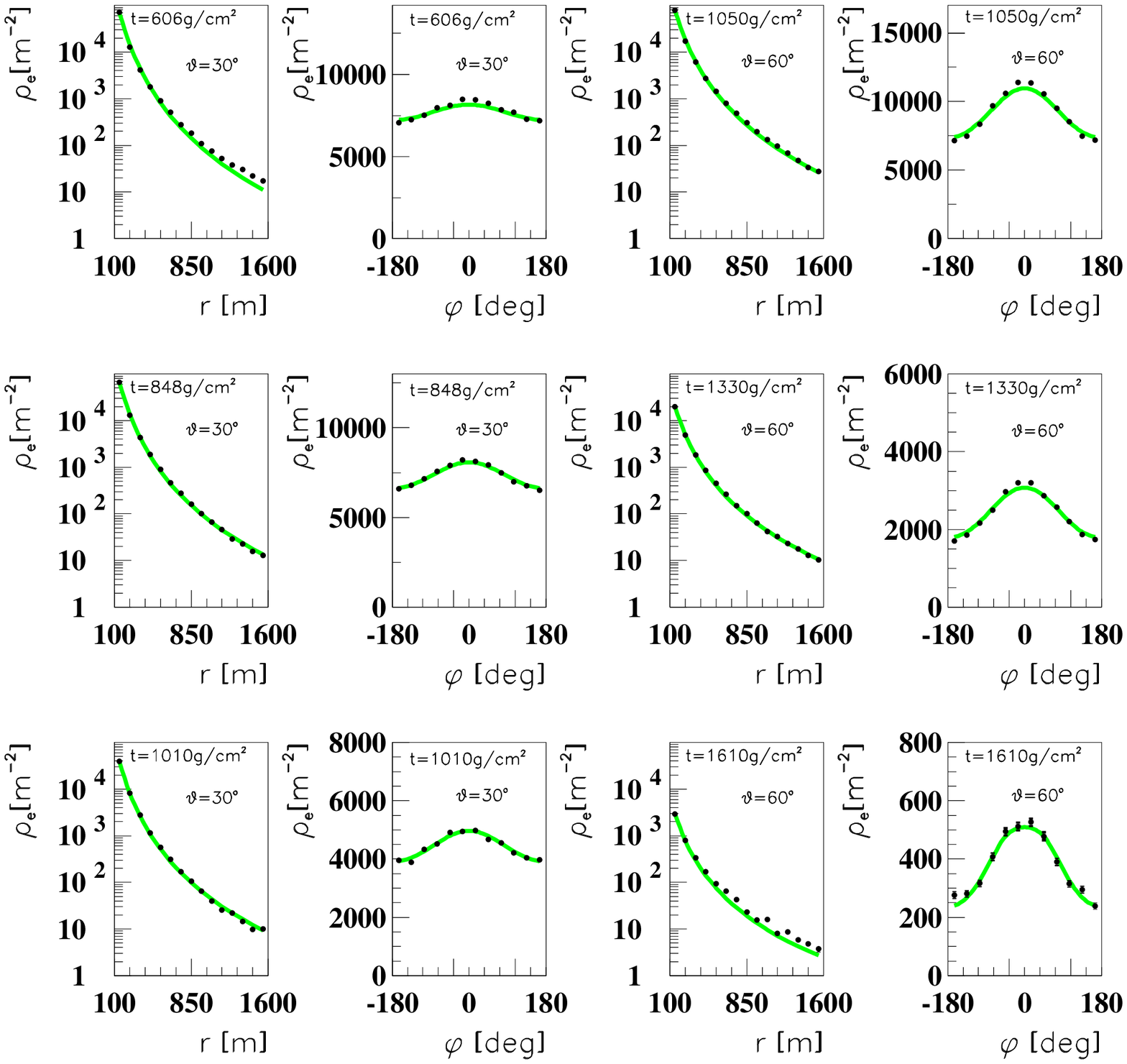,width=18.cm,clip=}
\caption{}
\end{center}
\end{figure}

Fig. 6: Projections of two-dimension electron distributions $r$-$\phi $ at 
different slant depths for a $10^{19}$eV proton shower at $30^{\circ}$
and $60^{\circ}$  zenith angles.

\newpage

\begin{figure}
\begin{center}
\epsfig{file=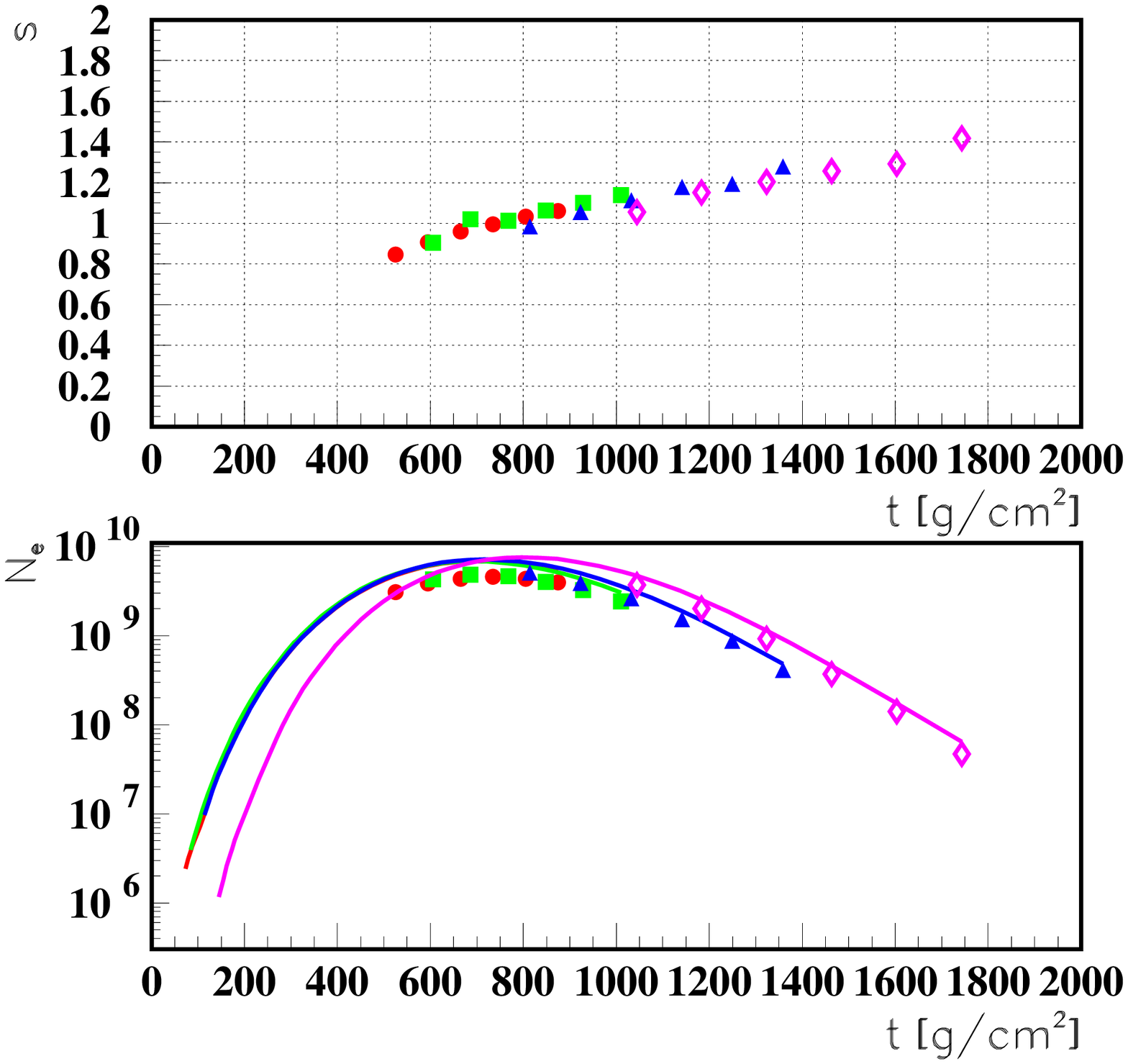,width=13.cm,clip=}
\caption{}
\end{center}
\end{figure}

Fig. 7: Upper plot: Age parameters at selected atmospheric depth 
 for showers entering at
 $0^{\circ}$ (circle),$30^{\circ}$(square), $50^{\circ}$ (triangle)
 and $60^{\circ}$ (diamond) zenith angles. Shower size obtained 
from the fits vs slant depth are shown in the lower plot. The actual 
total particle numbers for each single shower are indicated with solid 
lines.

\newpage
\begin{figure}
\begin{center}
\epsfig{file=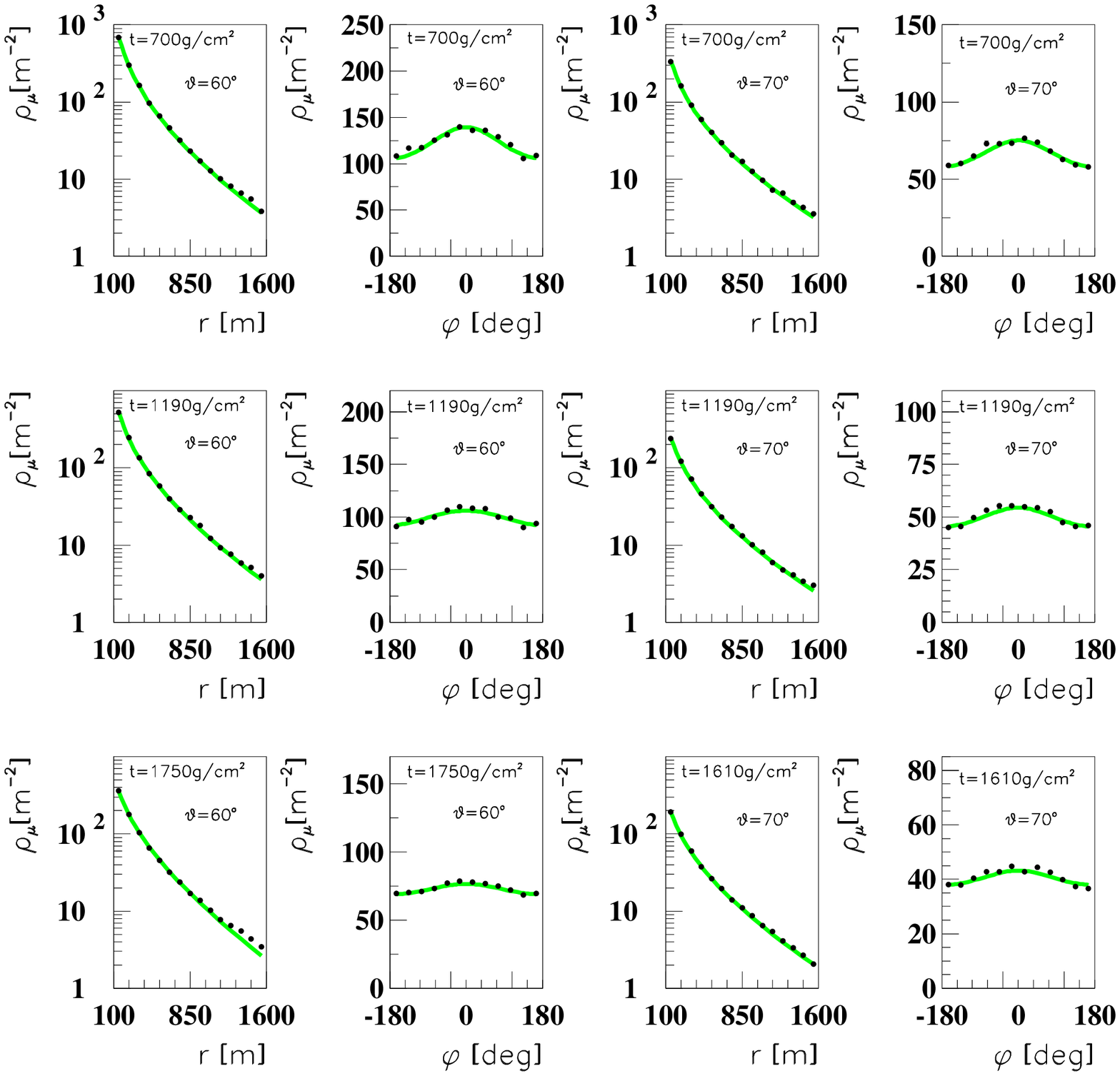,width=18.cm,clip=}
\caption{}
\end{center}
\end{figure}

Fig. 8: Projections of two-dimension muon distributions $r$-$\phi $ at 
different slant depths for a $10^{19}$eV proton shower at $60^{\circ}$ 
and $70^{\circ}$  zenith angles.

\newpage

\begin{figure}
\begin{center}
\epsfig{file=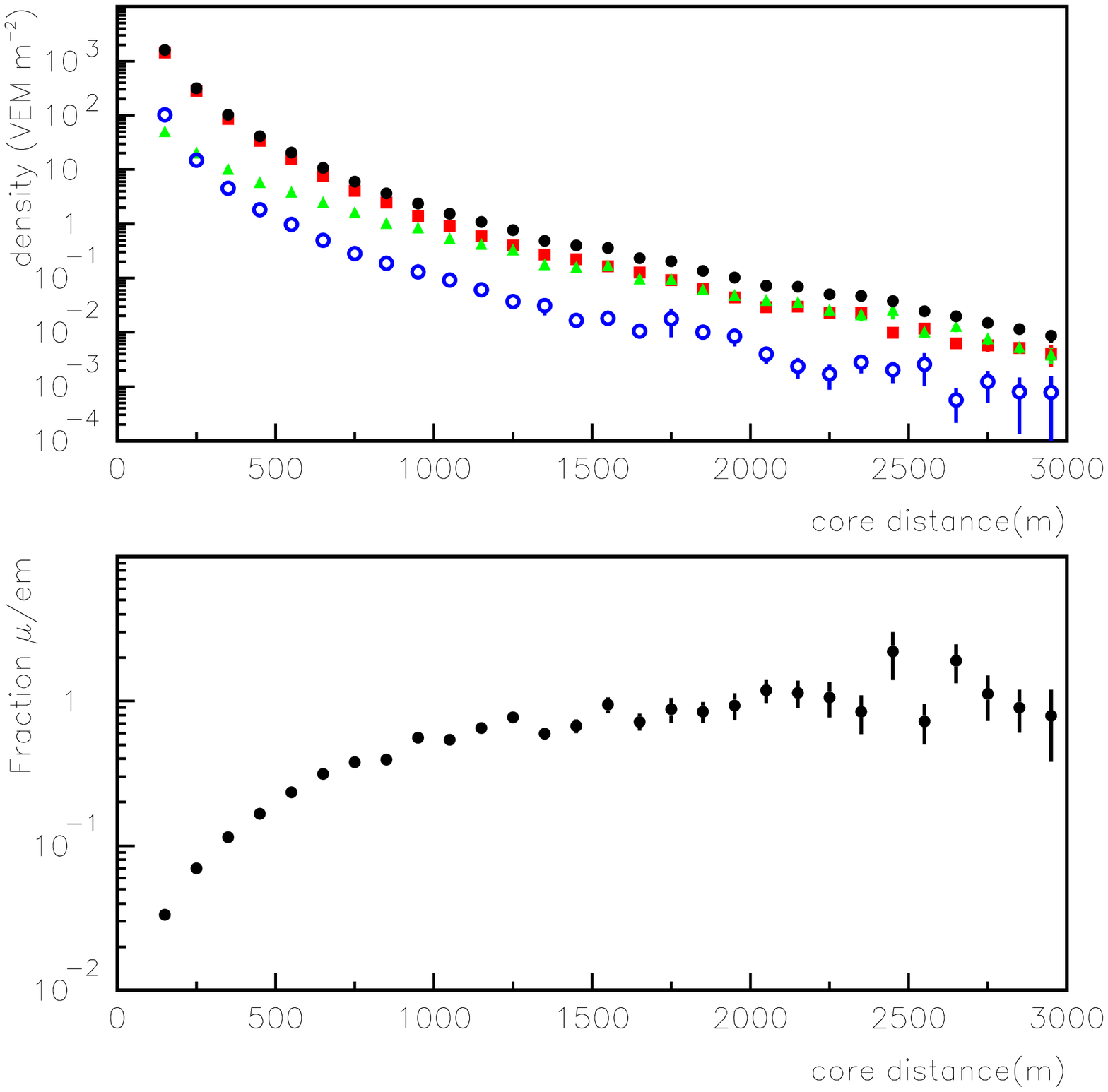,width=11.cm,clip=}
\caption{}
\end{center}
\end{figure}

Fig. 9: Top: Electron (open circles), gamma (squares), muon(triangle) 
and total (circles) water \v{C}erenkov signal in VEM m$^{-2}$ vs 
distance from the shower core for a $10^{19}$ proton vertical shower.
Bottom: Ratio of muon to 
electromagnetic signal vs distance from the core.

\newpage

\begin{figure}
\begin{center}
\epsfig{file=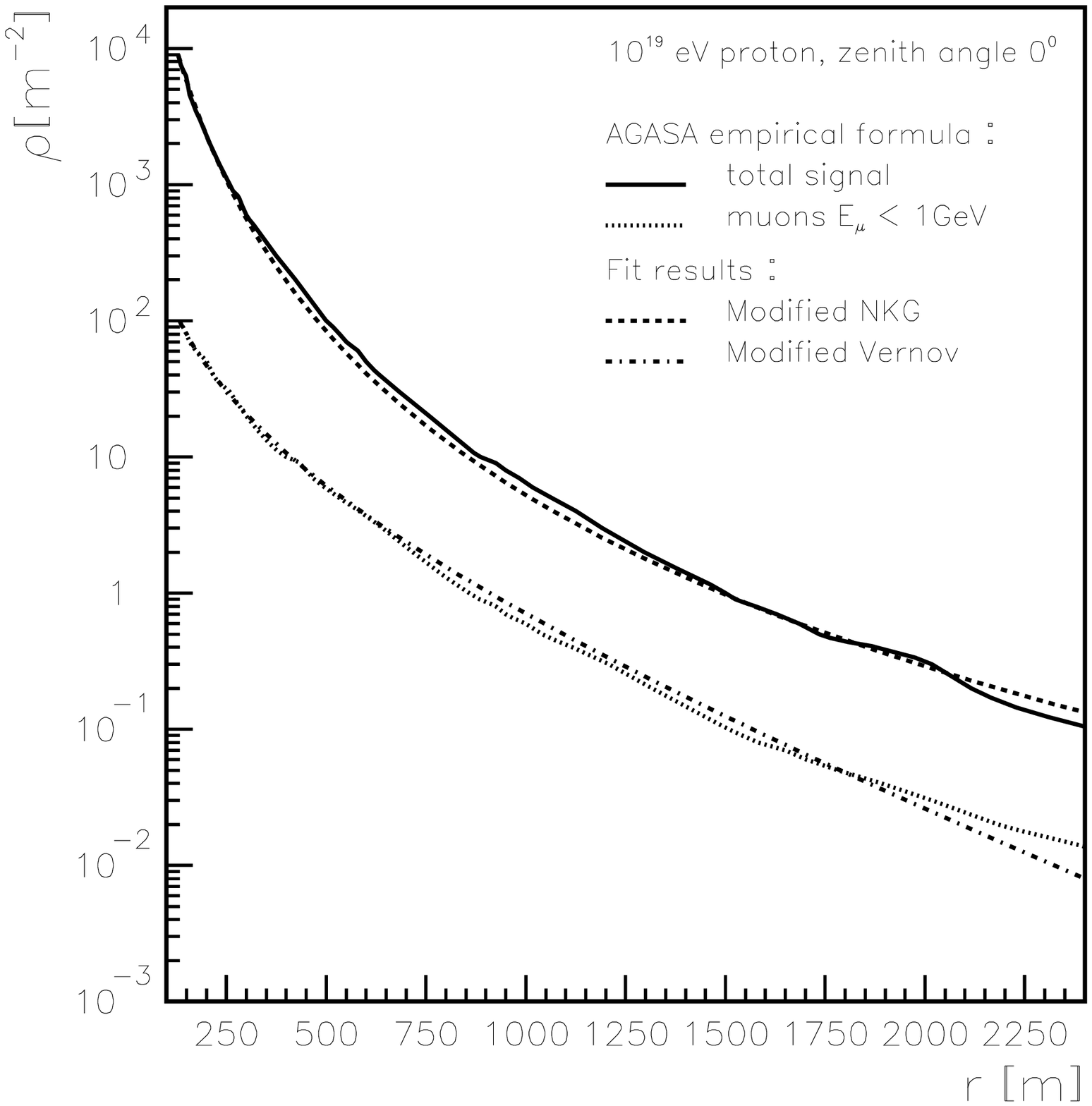,width=11.cm,clip=}
\caption{}
\end{center}
\end{figure}

Fig. 10: Lateral distributions of total signal (solid line) and muon(dotted line) determined by 
AGASA for vertical  $10^{19}$eV showers, compared with fits using 
modified NKG (dashed line) and Vernov (dash-dotted line) LDF. 

\newpage

\begin{figure}
\begin{center}
\epsfig{file=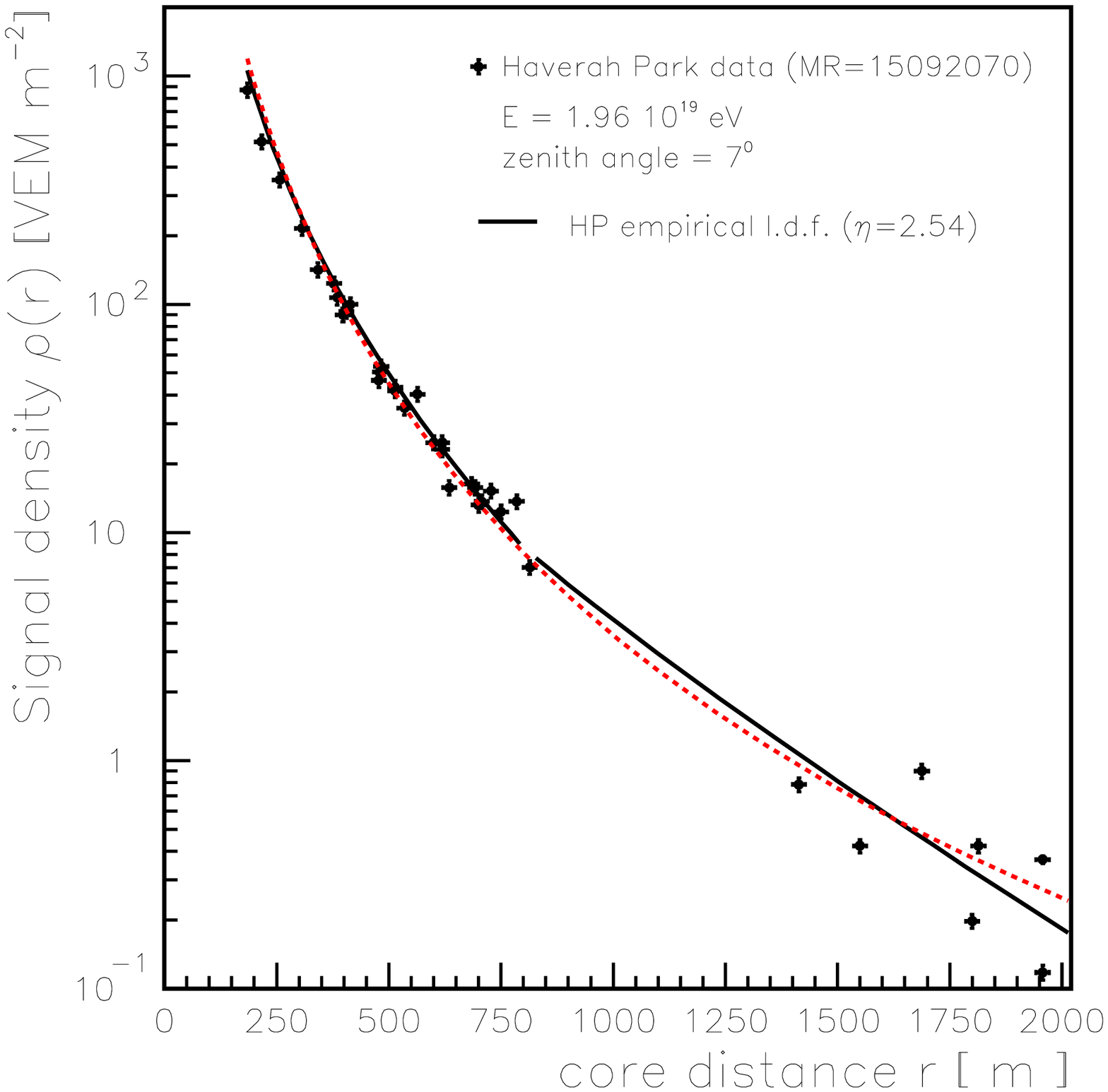,width=11.cm,clip=}
\caption{}
\end{center}
\end{figure}

Fig. 11: EAS observed by Haverah Park. The solid line indicates the Haverah 
Park predicted 
LDF and dashed-line corresponds to the result of the fit 
using the modified NKG LDF.

\newpage

\begin{figure}
\begin{center}
\epsfig{file=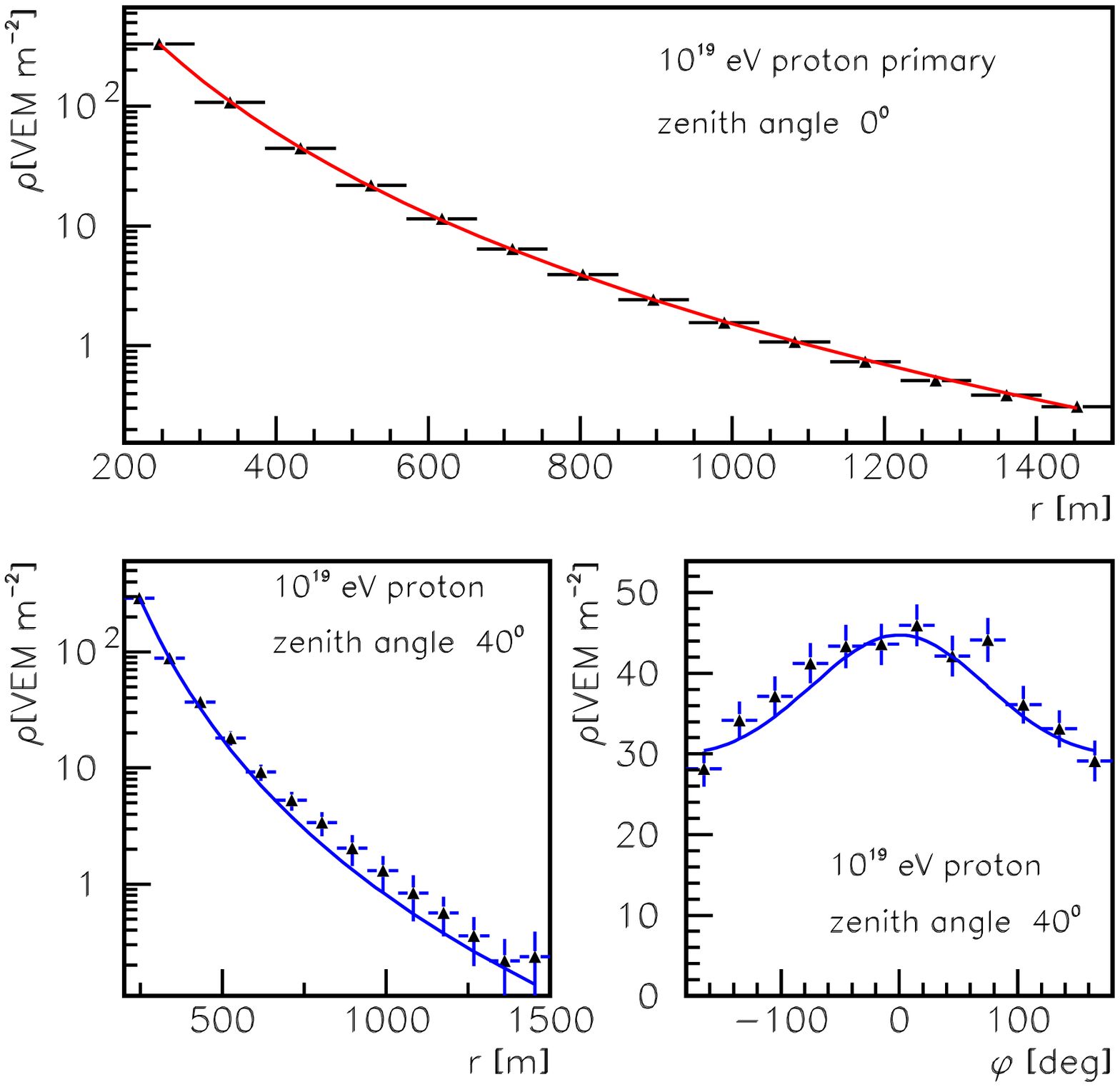,width=13.cm,clip=}
\caption{}
\end{center}
\end{figure}

Fig. 12: Top: Lateral distribution of signal produced by electrons, muons 
and photons in a water Cerenkov detector in units of vertical equivalent 
muons (see text). Bottom: Projections of two-dimension water Cerenkov signal distributions 
$r$-$\phi $ for a proton $40^{\circ}$. Solid  line corresponds 
to the fit performed usig the modified NKG LDF.


\begin{thebibliography}{00}

%
%

\bibitem{agasa}
N. Hasashida et al, Phys. Rev. Lett. 73 (1994) 3491.
\bibitem{hp}
M.A. Lawrence, R.J.O. Reid and A.A. Watson, J. Phys. G. 17 (1991) 733. 
\bibitem{yak}
D.J. Bird et al, Astrophys. J. 441 (1995) 144.
\bibitem{agasa2} 
N. Sakaki et al, AGASA Coll. Proc. 27th ICRC 2001:333.
\bibitem{hiresicrc}
A new cosmic ray spectrum as recorded by the HiRes Collaboration was presented
during 
the 27th ICRC, Hamburg, Germany (August 2001).
\bibitem{watsonagano}
M. Nagano and A.A. Watson, Rev. Mod. Phys. 72 (2000) 689.
\bibitem{zas}
M. Ave, R.A. Vazquez and E. Zas , Astropart. Phys.  14 (2000) 91.\\
M. Ave, J.A. Hinton, R.A. Vazquez, A.A. Watson and E. Zas, Astropart. Phys.  14 (2000) 109.
\bibitem{prlzas}
M. Ave, J.A. Hinton, R.A. Vazquez, A.A. Watson and E. Zas, Phys. Rev. Lett. 85,(2000) 2244.
\bibitem{desreport}
Auger Coll., Pierre Auger Project Design Report (2000) Fermi National Accelerator Laboratory.(www.auger.org/admin)
\bibitem{capelle}
K.S. Capelle et al, Astropart. Phys. 8 (1998) 321. 
\bibitem{xavierb}
X.Bertou et al, astro-ph/0104452.
\bibitem{hillas}
A.M. Hillas et al, Proc.  11th ICRC,Budapest (1969), Acta Physica Academiae Scietiarium Hungariae 29, Suppl. 3,(1970) 533.
\bibitem{andrews}
D. Andrews et al, Proc.  11th ICRC,Budapest (1969), Acta Physica Academiae Scietiarium Hungariae 29, Suppl. 3,(1970) 337.
\bibitem{antonov}
E.E. Antonov et al, JETP Letts. 69 (1999) 650.,JETP Lett. 68 (1998)185. 
\bibitem{ivanov}
A.A. Ivanov, V.P. Egorova, V.A. Kolosov, A.D. Krasilnikov, M.I. Pravdin and I.E. Sleptsov. , JETP Lett.69 (1999) 288.
\bibitem{cillis}
A.N. Cillis and S.J. Sciutto, Astropart. Phys. 14 (2001).
\bibitem{nkg}
K. Greisen, Ann. Rev. Nucl. Sci. 10 (1960) 63.
\bibitem{nk}
K. Kamata and J. Nishimura, Progr. Theor. Phys. Suppl 6 (1958) 93.
\bibitem{kascade}
T. Antoni et al, Astropart. Phys. 14 (2001) 245.
\bibitem{akeno}
M. Nagano, Y. Hatano, T. Hara, N. Hayashida, K. Kamata., T. Kifune, G. Tanahashi, andS. Kawaguchi, J. Phys. Soc. Jap. 53 (1984) 1667.
\bibitem{pdg}
Particle data group, Eur. Phys. J. C 15 (2000) 170.
\bibitem{hillas2}
A.M. Hillas, Proc. 16th ICRC, Tokyo, Japan, Vol.8 (1979) 7. \\
A.M. Hillas, Proc.  17th ICRC, Paris, France, Vol.8 (1981) 183. 
\bibitem{qgsjet}
N.N. Kalmykov and S.S. Ostapchenko, Phys. Atom. Nucl. 56 (1993) 346, Yad. Fiz. 56 (1993) 105.\\
N.N. Kalmykov , S.S. Ostapchenko and A.I. Pavlov, Bull. Russ. Acad. Sci. 58 (1994) 1966.
\bibitem{kascatere}
D. Heck; Kascade Coll., Proc. 27th ICRC, Hamburg (2001).
\bibitem{icrc01}
J. Milke; Kascade Coll., Proc. 27th ICRC, Hamburg (2001).
\bibitem{hillasthi}
A. M. Hillas, Nucl. Phys. B (Proc. Suppl.) 52B (1993) 29.\\
A. M. Hillas, Proc. 19th ICRC, La Jolla, Vol.1 (1985) 155.
\bibitem{aires}
S.J. Sciutto, astroph/9911331 (1999)  
\bibitem{linsley}
J. Linsley, Proc. 13th ICRC, Denver, Vol 5 (1973) 3212.  
\bibitem{aguirre}
C. Aguirre et al, Proc. 13th ICRC, Denver Vol 4 (1973) 2592.
\bibitem{porter}
N.A. Porter, Proc. 13th ICRC, Denver, Vol 5 (1973) 3657.  
\bibitem{kawaguchi}
S. Kawaguchi, K. Suga and H. Sakuyama, Proc. 14th ICRC, Munich, Vol 8 (1975) 2826.
\bibitem{akeno2}
M. Nagano et al, J. Phys. Soc. Jap. 53 (1984) 1667.
\bibitem{hara}
T. Hara et al, Proc. 16th ICRC Kyoto (1979). 
\bibitem{t5}
S. Yoshida et al, J. Phys. G: Nucl. Phys. 20 (1994) 651
\bibitem{t6}
A.V. Glushkov et al, Proc. 25th ICRC, Durban, Vol 6 (1997) 233. 
\bibitem{tt}
R.N. Coy et al., Astropart. Phys. 6 (1997) 263. 
\bibitem{capde}
M.F. Bourdeau, J. Capdevielle and J. Procureur, J. Phys. G: Nucl. Phys. 6 (1980) 901.
\bibitem{uchaikin}
A.V. Plyasheshnikov, A.A. Lagutin and V.V. Uchaikin, Proc. 16th ICRC Kyoto, Vol 7 (1979) 1. 
\bibitem{vernov}
S.N Vernov, G.B. Khristiansen, A.T. Abrasimov, V.B Atrashkevitch, I.F Beljaeva, G.V. Kulikov, K.V Mandritskaya, K.B Solovjeva, and B.A. Khrenov, Can. J. Phys 46 (1968) s197. 
\bibitem{alesio}
S. Alessio, H. Bilokon, B. D'Ettore Piazzoli, G. Mannochi, P. Pichi and K. Sitte, Nuovo Cimento 3 (1980) 573.
\bibitem{bosia}
G.F. Bosia, L. Briatore and K. Sitte, Nuovo Cimento 12 (1972) 1025.
\bibitem{berga}
L. Bergamasco, M. Castagnoli, M. Dardo,B. D'Ettore Piazzoli,G. Mannochi,P. Pichi, R. Visentin  and K. Sitte, Nuovo Cimento 34 (1976) 613.
\bibitem{pryke}
C. Pryke, Auger technical note GAP-98-034 (1998) (www.auger.org)
\bibitem{dova}
M.T Dova, L.N. Epele and A. Mariazzi, Proc. 26th ICRC Vol 1 (1999) 478.
\bibitem{nagano}
M. Nagano, D. Heck, K. Shinozaki, N. Inoue, 
J. Knapp, Astropart.Phys.13:277-294(2000).
\bibitem{sakaki}
N.Sakaki et al, Proc. 27th ICRC, 2001:380.
\bibitem{sergio}
S.Sciutto, private communication
\bibitem{clem}
C. Pryke, Pierre Auger technical note GAP-97-005 (1997), available elctronically at www.auger.org.
\bibitem{hpicrc}
M.Ave et al, Proc. 27th ICRC, 2001:381.
\end{thebibliography}
\end{document}